\documentclass[entropy,article,submit,moreauthors,pdftex,10pt,a4paper]{mdpi}

\firstpage{1} 
\articlenumber{x}
\doinum{10.3390/------}
\pubvolume{xx}
\pubyear{2016}
\copyrightyear{2016}
\externaleditor{Academic Editor: Martin Eckstein}
\history{Received: date; Accepted: date; Published: date}

\usepackage{amsmath}
\usepackage{amssymb}
\usepackage{graphicx}
\usepackage{grffile} 
\usepackage{color}
\usepackage{units}

\newcommand{\half}{\frac{1}{2}}

\newcommand{\kk}{\mathbf{k}}
\newcommand{\qq}{\mathbf{q}}
\newcommand{\pp}{\mathbf{p}}

\newcommand{\tave}{t_\mathrm{ave}}
\newcommand{\trel}{t_\mathrm{rel}}

\renewcommand{\Im}{\mathrm{Im}}

\newcommand{\ir}{\rceil}

\newif\ifLineNumbers \LineNumbersfalse
 \theoremstyle{mdpi}
 \newcounter{thm}
 \newcounter{ex}
 \newcounter{re}
 \theoremstyle{mdpidefinition}

\Title{Relationship between Population Dynamics and the Self-Energy in Driven Non-Equilibrium Systems}

\Author{Alexander F. Kemper$^{1,\ddagger}$* and James. K. Freericks$^{2,\ddagger}$}
\AuthorNames{Alexander F. Kemper and James. K. Freericks}

\address{%
$^{1}$ \quad Department of Physics, North Carolina State University; akemper@ncsu.edu\\
$^{2}$ \quad Department of Physics, Georgetown University; jkf@georgetown.edu}

\corres{Correspondence: akemper@ncsu.edu; Tel.: +1-919-515-7339}

\secondnote{These authors contributed equally to this work.}

\abstract{We compare the decay rates of excited populations directly calculated within a Keldysh formalism to the equation of motion of 
the population itself for a Hubbard-Holstein model in two dimensions. While it is true that these two approaches must give the same answer, 
it is common to make a number of simplifying assumptions within the differential equation for the populations that allows one to 
interpret the decay in terms of hot electrons interacting with a phonon bath. Here we show how care must be taken to ensure an 
accurate treatment of the equation of motion for the populations due to the fact that there are identities that require cancellations of 
terms that naively look like they contribute to the decay rates. In particular, the average time dependence of the Green's functions 
and self-energies plays a pivotal role in determining these decay rates.}

\keyword{population dynamics, non-equilibrium keldysh, scattering integrals}

\begin{document}
\nolinenumbers


%

\section{Introduction}


Non-equilibrium many-body physics is a complex problem because it requires the determination of two-time Green's functions 
within a system that does not have time-translation invariance. One pathway to simplify this approach has been to investigate the 
populations of electronic states (as a function of momentum and time) since they depend only on one time variable. 
Evaluating the differential equation of motion, which determines how the populations evolve with time, reveals that they are
complicated by ``memory effects'', given by the state of the system in the recent past, and hence they involve integrations over past times.
This non-locality in time arises entirely from the fact that we need to represent the two-particle averages (that typically give the potential 
energy of the system) via convolutions of the self-energy with the Green's function. One route to searching for a description that is local 
in time is to work on directly determining the two-particle Green's functions without resorting to this convolution. We do not pursue that 
approach here. Instead, we focus on examining the structure of the population dynamics and strive to understand as much as we can 
about the exact nature of these equations before we delve into approximate treatments. We also compare the approximate
results of the equations 
of motion for the populations with direct numerical results for the Green's functions and the self-energies that are found by self-consistently
solving the problem for the Green's functions within the Keldysh approach. This shows how the different scattering integrals behave as 
functions of time and how they ultimately determine the relaxation of the populations.

In recent years, non-equilibrium dynamical mean-field theory and closely related approaches\cite{freericks_nonequilibrium_2006,aoki_nonequilibrium_2014},
have solved numerous many-body physics
problems that are driven by external fields to model ultrafast pump-probe experiments
\cite{freericks_nonequilibrium_2006,freericks_theoretical_2009,moritz_electron-mediated_2013,sentef_examining_2013, kemper_effect_2014,
eckstein_photoinduced_2013,eckstein_ultrafast_2014,werner_role_2014,golez_dynamics_2015,eckstein_ultra-fast_2016}.
In those solutions, it is often seen that the populations
relax exponentially to their equilibrium values, and in some cases, one can identify the relaxation rate as given by the imaginary part of
the equilibrium self-energy. Furthermore, a simple linear-response analysis of the relaxation of a Green's function shows that the relaxation
rate is dominated by the imaginary part of the analytic continuation of the self-energy into the lower complex plane at the location of the pole
in the Green's function that lies closest to the real axis; if the system is described by a Fermi liquid, then the relaxation rate for electrons
lying near the Fermi energy is given essentially by the imaginary part of the self-energy at the excitation energy above the Fermi
energy\cite{galitskii1958translation}.

The fundamental question we address here is to determine the relationship between the relaxation rate of the populations and the
self-energy for more general non-equilibrium cases. One point, which makes the analysis complex, is that it is the average time
dependence that determines the relaxation of the system (because systems with no average time dependence have time translation
invariance, and do not evolve with time). But it is the relative time dependence which is most closely related to the frequency dependence
of the self-energy because the two are related via Fourier transformation. Reconciling these two issues is the primary goal of this work.
While we shed light onto this problem, we do not completely determine an analytic form for the population decay for the general
nonequilibrium case. 

\section{Results}


\subsection{Equations of motion}
\begin{figure}[h]
\centering
	\includegraphics[width=0.7\textwidth]{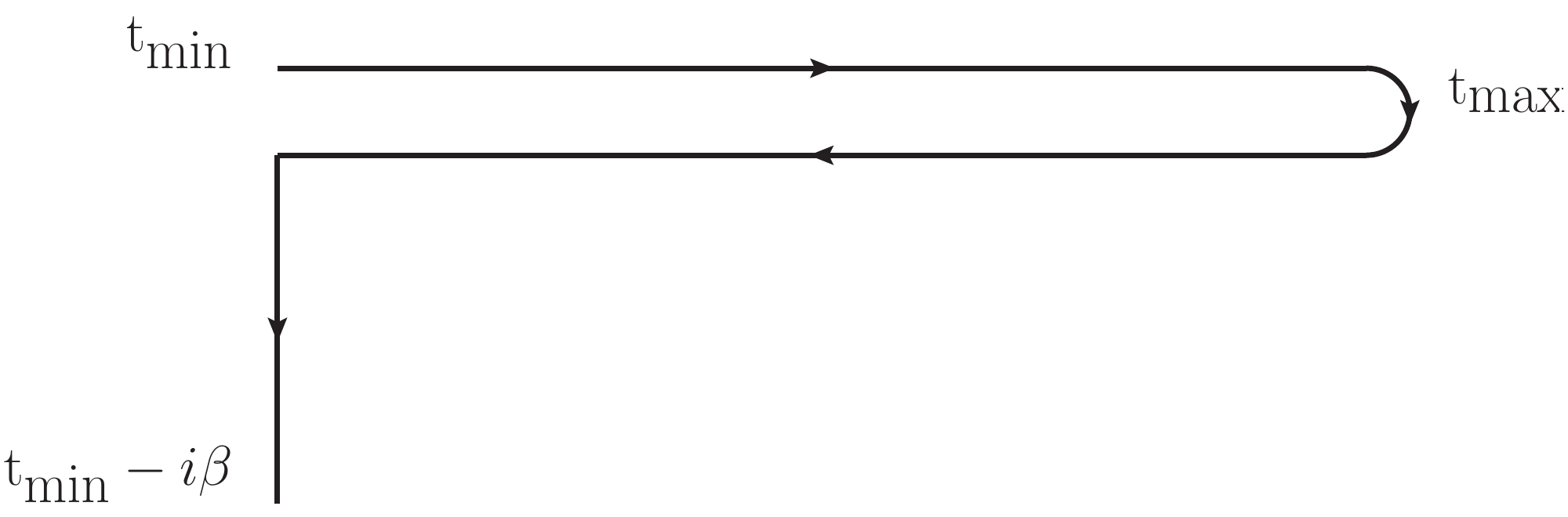}
	\caption{The Keldysh contour used in the analysis below. For convenience, $t_\mathrm{min}$ is set to equal $0$.}
	\label{fig:k_contour}
\end{figure}

The analysis of the two-time Green's function is done on the Keldysh contour, illustrated in Fig.~\ref{fig:k_contour}. The imaginary
spur allows for the calculation of an equilibrium thermal state at $t_\mathrm{min}$, which is subsequently propagated forward in time
along the contour using the equations of motion discussed below.
For convenience, we will assume $t_\mathrm{min}=0$.  The system is then subjected to an external perturbation (a field) at some
time $t>0$.
The Green's function for a state with quasi-momentum $\kk$ is defined on the
contour as

\begin{align}
G^\mathcal{C}_\kk(t,t') = -i \langle \mathcal{T_C} \hat c_\kk(t) \hat c^\dagger_\kk(t') \rangle,
\end{align}

\noindent where $\mathcal{T_C}$ is the contour time ordering operator, and $\hat c^\dagger_\kk / \hat c_\kk$ are the usual creation and annihilation
operators. The superscript $\mathcal{C}$ denotes a contour-ordered quantity. 
As discussed in Ref.~\cite{stefanucci_nonequilibrium_2013}, the Green's function
satisfies the equation of motion on the Keldysh contour:

\begin{align}
\left[ i\partial_t - \epsilon_\kk(t)\right] G^\mathcal{C}(t,t') = \delta^\mathcal{C}(t,t') + \int_\mathcal{C} \Sigma^\mathcal{C}(t,\bar t) G_\kk^\mathcal{C}(\bar t,t') d\bar t,
\end{align}

\noindent as well as a similar (adjoint) equation for $t'$. Here, the integration
is carried out over the entire contour.  $\Sigma$ encodes the interactions in a diagrammatic sense, and is a functional of
the Green's function.  For simplicity, we have chosen a momentum-independent self-energy.

Based on the location of the two times, we can break up the contour-ordered Green's function into different temporal components.  The real-time
Green's functions are the lesser, greater, retarded, and advanced components($<$, $>$, $R$, and $A$).  They are related through:

\begin{align}
G_\kk^<(t,t') &= G_\kk^\mathcal{C}(t \text{ on upper real branch}, t'\text{on lower real branch}), \\
G_\kk^>(t,t') &= G_\kk^\mathcal{C}(t \text{ on lower real branch}, t'\text{on upper real branch}), \\
G_{\kk}^R(t,t') &= \theta(t-t') \left[ G_{\kk}^>(t,t') - G_{\kk}^<(t,t') \right], \\
G_{\kk}^A(t,t') &= -\theta(t'-t) \left[ G_{\kk}^>(t,t') - G_{\kk}^<(t,t') \right].
\end{align}

\noindent The mixed components, when one of the arguments ($\tau$) is on the imaginary spur, 
and the other ($t$) on the real time branch, are denoted
as $G^\rceil(t,\tau)$ and $G^\lceil(\tau,t)$.
Finally, the component which has both arguments on the imaginary spur is the usual Matsubara Green's function $G^M(\tau,\tau')$.
These subdivisions are applied for all contour-ordered quantities.

We can carry this through for the equation of motion, which splits into 3 distinct pieces, and similarly for the adjoint equation.
For completeness, we list them here, following the notation of  Ref.~\cite{stefanucci_nonequilibrium_2013}.
Since time-translation invariance applies on the imaginary axis, the equation of motion is significantly simplified:

\begin{align}
\left[-\partial_\tau - \epsilon_\kk(0)\right] G^M(\tau) = \delta(\tau) + \int_0^\beta \Sigma^M(\tau-\bar\tau) G^M_\kk(\bar \tau).
\end{align}

\noindent The lack of dependence on the real time integrations is a reflection of causality \textemdash the equilibrium state on the
imaginary spur should not depend on any quantities that occur later in time.
The equations involving real times are:

\begin{align}
\left[ i\overrightarrow{\partial_t} - \epsilon_\kk(t)\right] G_{\kk}^\gtrless(t,t') &= I^\gtrless_{1,\kk}(t,t'), \label{eq:dtgless1}\\
G_{\kk}^\gtrless(t,t') \left[ -i\overleftarrow{\partial_{t'}} - \epsilon_\kk(t')\right] &= I^\gtrless_{2,\kk} (t,t'), \label{eq:dtgless2}\\
\left[ i\overrightarrow{\partial_t} - \epsilon_\kk(t)\right] G_{\kk}^\rceil(t,\tau) &= I^\rceil_{1,\kk}(t,\tau), \\
G_{\kk}^\lceil(\tau,t) \left[ -i\overleftarrow{\partial_{t}} - \epsilon_\kk(t')\right] &= I^\lceil_{2,\kk}(\tau,t) ,
\end{align}

\noindent where the derivative applies following the direction of the arrow.  The terms on the right hand side are the scattering integrals,
which we will focus on:

\begin{align}
I^{\gtrless}_{1,\kk}(t,t') &= \int_0^t d\bar t \Sigma^R(t,\bar t) G^{\gtrless}_\kk(\bar t,t)  + \int_0^{t'} d\bar t\Sigma^\gtrless(t,\bar t) G_\kk^A(\bar t,t)+ \frac{1}{i} \int_0^\beta d\bar\tau\ \Sigma^{\rceil}(t,\bar \tau) G^{\lceil}_\kk(\bar \tau, t) , \label{eq:I1gtrless}\\
I^{\gtrless}_{2,\kk}(t,t') &= \int_0^t d\bar t G_\kk^R(t,\bar t) \Sigma^{\gtrless}(\bar t,t)  + \int_0^{t'} d\bar t G_\kk^\gtrless(t,\bar t) \Sigma^A(\bar t,t) + \frac{1}{i} \int_0^\beta d\bar\tau\ G_\kk^{\rceil}(t,\bar \tau) \Sigma^{\lceil}(\bar \tau, t) , \\
I_\kk^\lceil(t,\tau) &= \int_0^t d\bar t \Sigma^R(t,\bar t) G^\lceil(\bar t,\tau) + \frac{1}{i} \int_0^\beta d\bar\tau \Sigma^\lceil(t,\bar \tau) G_\kk^M(\bar\tau -\tau) ,\\
I_\kk^\ir(\tau,t) &= \int_0^t d\bar t G^\ir(\tau,\bar t) \Sigma^A(\bar t, t) + \frac{1}{i} \int_0^\beta d\bar\tau G^M_\kk(\tau-\bar\tau) \Sigma^\ir(\bar\tau,t).
\end{align} 

\noindent The scattering integrals have two distinct sets of terms: those involving values on the imaginary axis, and those
that do not.  In the presence of any type of interactions, the Green's functions decay in the $t-t'$ direction. 
This leads to a decay in the contribution of the mixed pieces
to the integrals as time gets further from $t=0$.  We will assume that $t$ and $t'$ have advanced sufficiently far from $t=t'=0$ that we can neglect
those mixed pieces in our analysis.

The density for momentum $\kk$ is given by the time-diagonal piece of the lesser Green's function, $n_\kk(t) = -i G_\kk^<(t,t)$.
To propagate along the $t=t'$ ($\tave$) direction, we need to combine Eqs.~(\ref{eq:dtgless1}) and (\ref{eq:dtgless2}). This yields

\begin{align}
i\partial_t G_\kk^<(t,t) &= \left[ \epsilon_\kk(t), G_\kk^<(t,t) \right] + I^<_{1,\kk}(t,t) - I_{2,\kk}^<(t,t),
\end{align}

\noindent which can be simplified by noting that $I^\gtrless_{1,\kk}(t,t') = -I^\gtrless_{2,\kk}(t',t)^\dagger$.
For single-band models on lattices without a basis the Green's function can be described by a set of scalars
and the energy $\epsilon_\kk(t)$ commutes with the Green's function. We can further assume that we have gone far
enough along in $t$ and $t'$ that we can extend the lower bound of the integral to $-\infty$ due to the Green's function decay, and
are left with

\begin{align}
i\partial_t G_\kk^<(t,t) = \int_{-\infty}^t d\bar t &\left\lbrace \Sigma^R(t,\bar t) G_\kk^<(\bar t,t) + \Sigma^<(t,\bar t) G^A_\kk(\bar t, t) \right. \nonumber\\
& \left. - G^<_\kk(t,\bar t) \Sigma^A(\bar t, t) - G_\kk^R(t,\bar t) \Sigma^<(\bar t, t) \right\rbrace, \label{eq:rhs} \\
\equiv \Sigma^R \cdot G^<_\kk &+ \Sigma^< \cdot G^A_\kk - G^<_\kk \cdot \Sigma^A - G^R_\kk \cdot \Sigma^< \label{eq:rhs_v1},
\end{align}

\noindent where in the last line we have introduced the $\cdot$ (as a non-Abelian operator)
to represent the integral for notational convenience.

\subsection{General remarks on the scattering integrals}

It is instructive to consider the contour-ordered quantities as functions of the average and relative time, rather than the explicit
$t$ and $t'$: $\tave = \half \left(t+t'\right), \trel = t-t'$.
The population dynamics occurs along the $\tave$ direction, whereas quasiparticle lifetimes arise from the $\trel$ direction.
The scattering integrals control the temporal dynamics of the Green's function along the $\tave$ direction.  Thus, an examination of
the integrals can lead to some general insights regarding the dynamics.
First, it can be shown that an average time dependence is necessary for any change to occur \textemdash that is, without any
dependence on $\tave$, the scattering integrals (and thus the time derivative of the density) are 0. 

It is illustrative to consider the equations in a form where the Fourier transform over $\trel$ has been performed.
For this transform to be well-defined, we must
assume that any average time
dependence is slow enough within the window of Fourier transformation set by the decay of the Green's function along $\trel$
that we can replace the average time dependence with a single time only, namely $\tave$.
Any applied field is assumed to have occurred sufficiently far in the past that it falls outside this window.  This yields

\begin{align}
i\partial_{\tave} G_\kk^<(\tave, \omega) = &2i\Im\left[\Sigma^R(\tave, \omega)\right] G^<(\tave, \omega)  \nonumber \\
-& 2i\Im\left[G^R_\kk(\tave, \omega)\right] \Sigma^<(\tave, \omega).
\label{eq:RHS_tave_omega}
\end{align}

\noindent When there is no dependence on $\tave$, we can use Dyson's equation and substitute for $G_\kk^<(\omega)$ and $\Im G_\kk^R(\omega)$.
Long after $t=t'=0$ where the mixed components have decayed, we can re-write Dyson's equation [Eqs.~(\ref{eq:dtgless1}) and (\ref{eq:I1gtrless})] in the
following way:

\begin{align} 
G_\kk^< &= G_\kk^R \cdot \Sigma^< \cdot G^A_\kk.
\end{align}

\noindent Making the same transformation to $\tave,\trel$ and Fourier transforming over $\trel$, we can substitute this result into Eq.~(\ref{eq:RHS_tave_omega}):

\begin{align}
i\partial_{\tave} G^<(\tave, \omega) = &2i\Im\left[\Sigma^R(\omega,\tave)\right] \Sigma^<(\omega,\tave) |G_\kk^R(\omega,\tave)|^2  \nonumber \\
 -&2i\Im\left[\Sigma^R(\omega,\tave)\right] |G^R_\kk(\omega,\tave)|^2 \Sigma^<(\omega,\tave) \\
&= 0.
\end{align}

\noindent The point here is that the lack of time dependence in the density comes from the cancellation of two terms, rather than each term
being 0 individually.  This will come up again when we consider specific cases. 

The central result is even stronger than this: once the lesser Green's function can be represented by any distribution function of average time and frequency
that multiplies the imaginary part of the retarded Green's function at the same average time and frequency, that is 
$G_\kk^<(\tave,\omega) = -2i f_\text{dist}(\omega,\tave) \Im G_{\kk}^R(\omega,\tave)$ (and similar for the self-energy), 
then the population no longer changes with
time.
The factor of two arises from the connection
to equilibrium; this assumption is true if the system has
equilibriated at an elevated temperature $T_{el}$, where the fluctuation-dissipation theorem holds and $f(\omega,T_{el})$ is the Fermi function.
 We can readily see this from Eq.~(\ref{eq:RHS_tave_omega}) by substituting this relation:
 
\begin{align}
i\partial_{\tave} G^<(\tave, \omega) = &4 \Im\left[\Sigma^R(\omega,\tave)\right]  f_\text{dist}(\omega,\tave) \Im G_{\kk}^R(\omega,\tave) \nonumber \\
 -&4\Im\left[G^R(\omega,\tave)\right] f_\text{dist}(\omega,\tave)\Im \Sigma^R(\omega,\tave) \\
&= 0.
\end{align}

\noindent Note that this cancellation occurs for all types of scattering and for all strengths of the scattering. 
It does not require any so-called "bottlenecks" but simply follows from the dynamics of the populations and the exact form of the scattering integral.
A further implication of this cancellation is that a simple hot electron model cannot fully describe the dynamics 
because approximating $G^<_\kk(\omega,\tave)$ and $\Sigma^<(\omega,\tave)$ by the retarded components multiplied by the \textit{same} average-time
dependent distribution function
will not 
work, because such systems do not evolve in time any further, even though the distributions are different from the equilibrium distributions.


\subsubsection{Self-consistency}

A point should be raised here regarding self-consistency \textemdash that is, using the density after the pump
to determine the self-energy, rather than using the equilibrium density. This is equivalent to using the unrenormalized Green's function
to evaluate the self-energy, which is often done in equilibrium approaches to evaluate first-order effects.  As we will show, this is not
the same in the time domain.
Without self-consistency, the above approach cannot be applied in the same
fashion because the relation between $\Sigma$ and $G_\kk$ is not as clear.  
Instead, as above, let us assume that some time after the pump, the system can be described as a population occupying 
an equilibrium spectral function
\textemdash that is, $G_\kk^<(\tave,\omega) =-2i f_\text{dist}(\omega,\tave) \Im G_{\kk,\text{eq}}^R(\omega)$.  Then, if the self-energy is given by the
original \textit{equilibrium} one, we find

\begin{align}
i\partial_{\tave} G^<(\tave, \omega) = &4 \Im\left[\Sigma^R(\omega)\right] \Im \left[G_{\kk,\text{eq}}^R(\omega)\right]\left[ f_\text{dist}(\omega,\tave) - f_\text{eq}(\omega) \right],
\label{eq:RHS_temps}
\end{align}

\noindent where $f_\text{eq}(\omega)$ is the equilibrium distribution used to evaluate self-energy.
%
%
If the self-energy is thus determined based on the equilibrium Green's functions, there cannot be a balance of terms
and the density will decay until it reaches the initial equilibrium state, independent of the types of interactions present. 
More generally, if the distribution is the same for $G_\kk^<(\tave,\omega)$ as for $\Sigma^<(\omega)$, then there is no further relaxation.

One way this can be further understood is by breaking the self-energy 
into ``dark'' and ``induced'' pieces,
following the work by \u{S}pi\u{c}ka \textit{et al.}\cite{spicka_long_2005,spicka_long_2005-1,spicka_long_2005-2}.  Starting from Eq.~(\ref{eq:rhs_v1}) we
find 

\begin{align}
i\partial_{\tave} G_\kk^<(\tave, \omega) = \Sigma^R_D \cdot G_\kk^< - \Sigma^<_D \cdot G_\kk^A + \Sigma^R_I \cdot G_\kk^< - \Sigma^<_I \cdot G_\kk^A
+ \mathrm{H.C.}
\end{align}

\noindent At this point, we can identify the first two terms on the
RHS as those captured by approaches lacking self-consistency \textemdash they capture the changes in the Green's function,
but not the self-energy.  If we limit the RHS to these terms and apply the approach discussed above, it becomes clear
how the population dynamics have a simple connection to the \textit{equilibrium} (or ``dark'') retarded self-energy:

\begin{align}
i\partial_{\tave} G_\kk^<(\tave, \omega) &= 4 \Im\left[\Sigma_D^R(\omega) \right] \Im\left[G_{\kk,D}^R(\omega)\right] \left[ f(\omega,\tave) - f_{eq}(\omega) \right] \label{eq:RHS_temps0}
\\
&= -2i \Im\left[\Sigma_D^R(\omega) \right]  G^<_{\kk,I}(\tave,\omega) .
\end{align}

\noindent As long as the induced density is different from 0, we can observe a decay with an
exponential decay constant $\tau(\omega)^{-1} = -2\Im\Sigma_D^R(\omega)$.
This is the relation between the population decay and self-energy reported previously\cite{kemper_mapping_2013,sentef_examining_2013}.  It is the same
relation as that between the decay of the Green's function in \textit{relative time $\trel$}, but does not have the same origin.  In that case,
the self-energy gives rise to a line width in the spectrum of a single excitation, as measured e.g. by equilibrium photoemission spectroscopy.
Here, the relation arrives only through copious approximations applied to population dynamics\cite{spicka_long_2005,spicka_long_2005-1,spicka_long_2005-2},
which occur along an orthogonal
direction to $\trel$. We require the population dynamics to be very slow for this to hold, or for the pump-induced population
changes to be extremely small such that only these terms play a role.  Once the approximations are violated, more complex dynamics
arise, as discussed by several authors.\cite{stefanucci_nonequilibrium_2013,marini_competition_2013,kemper_effect_2014,murakami_interaction_2015}  This can already be seen from Eq.~(\ref{eq:RHS_temps}),
which bears a remarkable similarity to Eq.~(\ref{eq:RHS_temps0}) except that the full pump-modified self-energy has to be
accounted for. This leads to modification of the observed interactions by the pump, as discussed in.\cite{kemper_effect_2014,rameau_photoinduced_2014,ishida_quasi-particles_2016} Evidently, the lack of self-consistency is an uncontrolled
approximation in the time domain, which can fail to capture important aspects of the population dynamics.

\subsubsection{Specific examples}

As a concrete illustration of the above, let us consider a simple form of impurity scattering which
can account for the changes in the density due to the pump: the self-consistent
Born approximation, where the self-energy is proportional to the local Green's function: $\Sigma^R(t,t') = V^2 \sum_\kk G_\kk^R(t,t')$,
and similarly for the other Keldysh components.  Inserting this into Eq.~(\ref{eq:rhs_v1}), and applying the equations above,
we find

\begin{align}
i\partial_{\tave} G_\kk^<(t,t) &= V^2 \sum_\pp \left[ G^R_\pp \cdot G^<_\kk + G^<_\pp \cdot G^A_\kk -
G^<_\kk \cdot G^A_\pp - G^R_\kk \cdot G^<_\pp \right]  \\
i\partial_{\tave} G^<_\kk(\tave,\omega) &= 2iV^2 \sum_\pp \left\lbrace
\Im\left[G^R_\pp(\omega,\tave)\right] G^<_\kk(\omega,\tave) - \Im\left[G^R_\kk(\omega,\tave)\right] G^<_\pp(\omega,\tave) 
\right\rbrace \\
&\approx 2iV^2 \sum_\pp 
\Im\left[G^R_\pp(\omega)\right] \Im\left[G^R_\kk(\omega)\right]  \left\lbrace f(\omega,\tave) -  f(\omega,\tave)  \right\rbrace\\
&= 0.
\end{align}

\noindent This is a particularly striking result in light of the usual equilibrium connection between the quasiparticle lifetime and the self-energy.
Here, the self-energy is clearly finite, whether calculated based on the pumped states or the equilibrium states, yet the population
decay rate is 0 once the system has reached a state where the distribution function is independent of momentum.

As a final note, we may apply the above methodology to a system of purely interacting electrons. The lowest-order term
that leads to scattering of more than 1 electron is at 2$^\mathrm{nd}$ order, with

\begin{align}
\Sigma_\kk(t,t') = \sum_{\qq\pp} U_{\qq} G_{\kk-\qq}(t,t') G_\pp(t,t') G_{\pp+\qq}(t,t').
\end{align}

\noindent Although the algebra is more complex, it can nevertheless be shown that once the electrons have been approximated as a
thermal distribution occupying the spectral function,
the RHS of the equations of motion are identically 0.  This is as expected, since a thermalized state is the stable solution
for a system of interacting electrons.  Indeed, the specific form of the scattering mechanism is not important, the dependence of
the lesser quantities on the distribution function is the important element.

\subsection{Numerical evaluation of the scattering integrals}

To evaluate the approximations and confirm the arguments made above, we perform numerical evaluation of the equations
of motion on the Keldysh contour. The method has been described in some detail in Refs.\cite{stefanucci_nonequilibrium_2013,kemper_effect_2014}.
We will follow the same separation used and motivated in the previous section, and consider separately the terms
$I^\Sigma _\kk\equiv \Sigma^R \cdot G^<_\kk - G^<_\kk \cdot \Sigma^A$ and 
$I^G _\kk \equiv G^R_\kk \cdot \Sigma^< - \Sigma^< \cdot G_\kk^A$. 
These are different from $I^1_\kk$ and $I^2_\kk$, but rather are the full scattering integral separated into the
terms that lead to two terms as they appear in Eq.~(\ref{eq:RHS_tave_omega}).
As noted above, these two terms should balance when there is no further time dependence to the density.
The full
scattering integral is then a difference between the two [see Eq.~(\ref{eq:rhs_v1})].  

As an illustrative model, we choose the Holstein-Hubbard Hamiltonian with local scattering:

\begin{align}
\mathcal{H} = \sum_{\kk,\sigma} \epsilon_\kk \hat c^\dagger_{\kk,\sigma} \hat c_{\kk\sigma} + U \sum_i \hat n_{i\uparrow} \hat n_{i\downarrow} 
+ g \sum_{k,q,\sigma} \hat c^\dagger_{k+q,\sigma} \hat c_{k,\sigma} \left( \hat b_q + \hat b^\dagger_{-q} \right) + \Omega \sum_i 
\left( \hat b^\dagger_i \hat b_i + \half\right)
,\label{eq:hamiltonian}
\end{align}

\noindent where $\hat c^\dagger / \hat c$ and $\hat b^\dagger / \hat b$ are the creation and annihilation operators for electrons and phonons, respectively,
and $\hat n_i \equiv \hat c^\dagger_i \hat c_i$ is the local density operator.

\begin{figure}[h]
	\includegraphics[width=\textwidth]{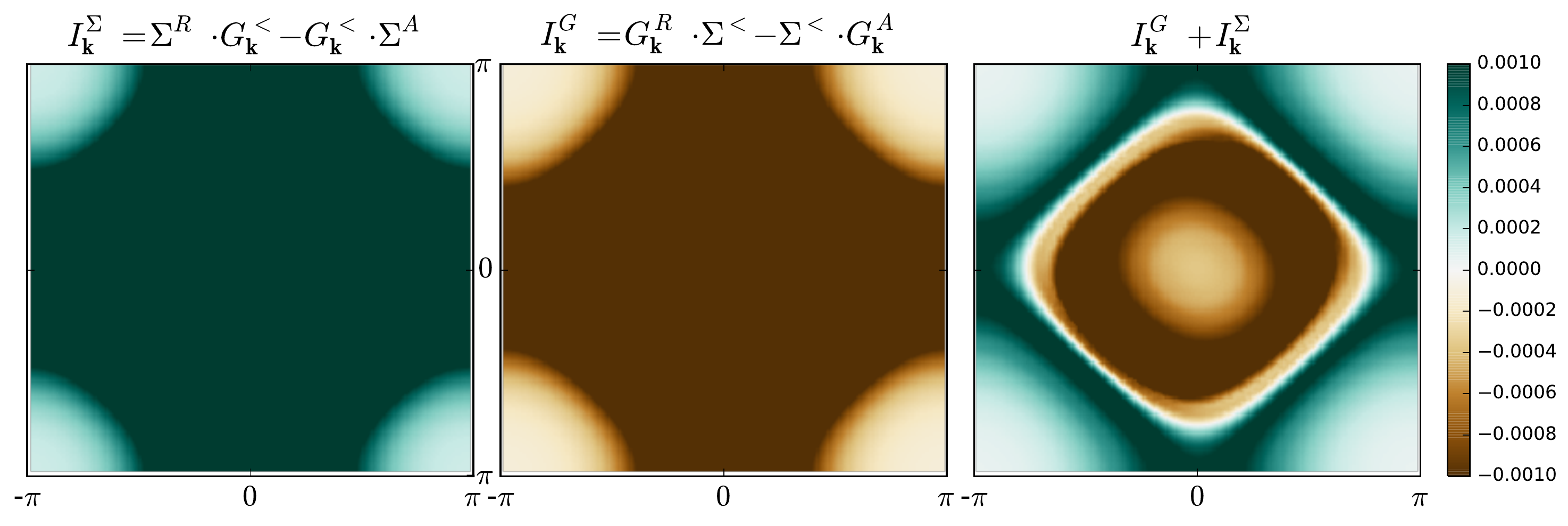}
	\includegraphics[width=\textwidth]{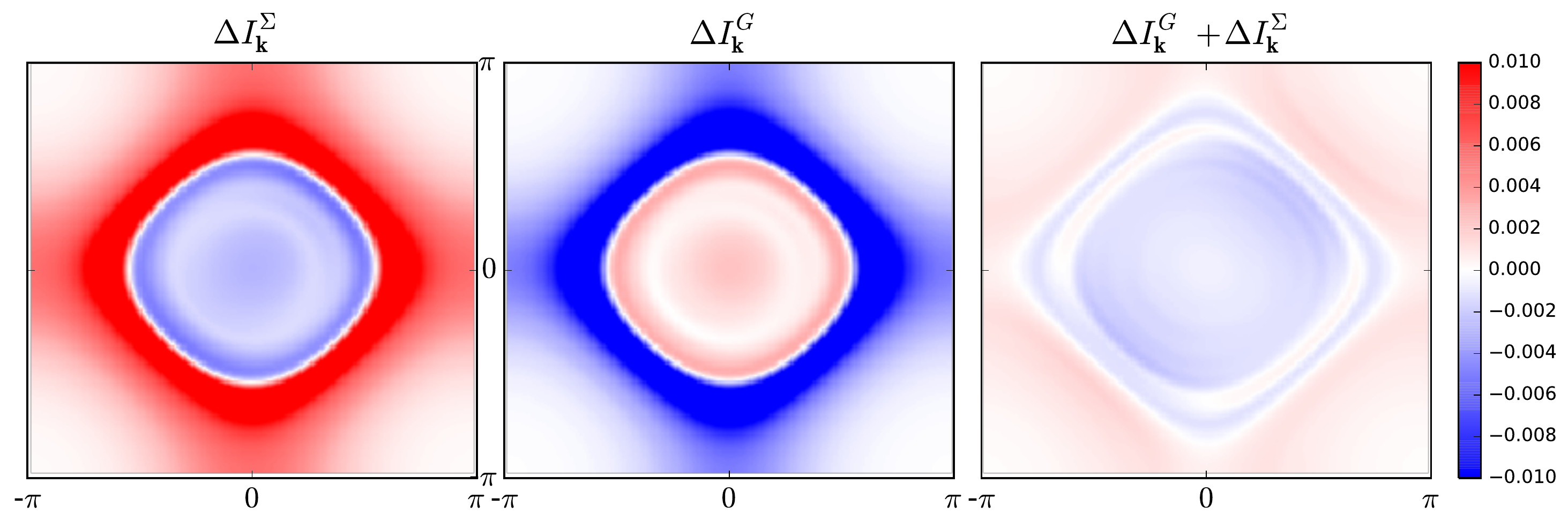}
	\caption{Field-induced changes in the momentum dependent scattering integrals, where $t=100$ eV$^{-1}$.
	The panels show the two individual contributions as well as the difference. Top: absolute values. Bottom: difference
	from equilibrium.
	}
	\label{fig:rhs_snapshots}
\end{figure}

The electron-electron interactions are treated at second order,
with 

\begin{align}
\Sigma_\mathrm{e-e}^\mathcal{C}(t,t') = U^2 G^\mathcal{C}_\mathrm{local}(t,t') \star G^\mathcal{C}_\mathrm{local}(t,t') \star G^\mathcal{C}_\mathrm{local}(t',t)
\end{align}

\noindent where the star ($\star$) denotes a \textit{product} on the Keldysh contour. When present,
$U^2=0.09$ eV$^2$. 
The electron-phonon scattering is treated at first order
within the Migdal-Eliashberg formalism, with

\begin{align}
\Sigma_\mathrm{e-p}^\mathcal{C}(t,t') = g^2 G^\mathcal{C}_\mathrm{local}(t,t') \star D_0^\mathcal{C}(t,t'),
\end{align}

\noindent where $D_0^\mathcal{C}(t,t')$ is the propagator for a bare Einstein phonon with frequency $\Omega$.
When present, $g^2=0.01$ eV$^2$ and $\Omega=0.2$ eV.
We drive with two-dimensional tight-binding model with
nearest neighbor hopping $V_\mathrm{nn}=0.25$ eV and a chemical potential $\mu=-0.2$ eV.
The field is applied using minimal coupling in the Hamiltonian gauge. We use a simple oscillatory field in the zone diagonal direction
with $A(t) = A_\mathrm{max} \sin(\omega t) \exp(-\half (t-t_0)^2 / \sigma^2)$, where $\omega=0.5$ eV, $t_0=60$ eV$^{-1}$, $\sigma=10$ eV$^{-1}$
and $A_\mathrm{max} = 0.8$.

\begin{figure}[h]
\centering
	\includegraphics[width=\textwidth]{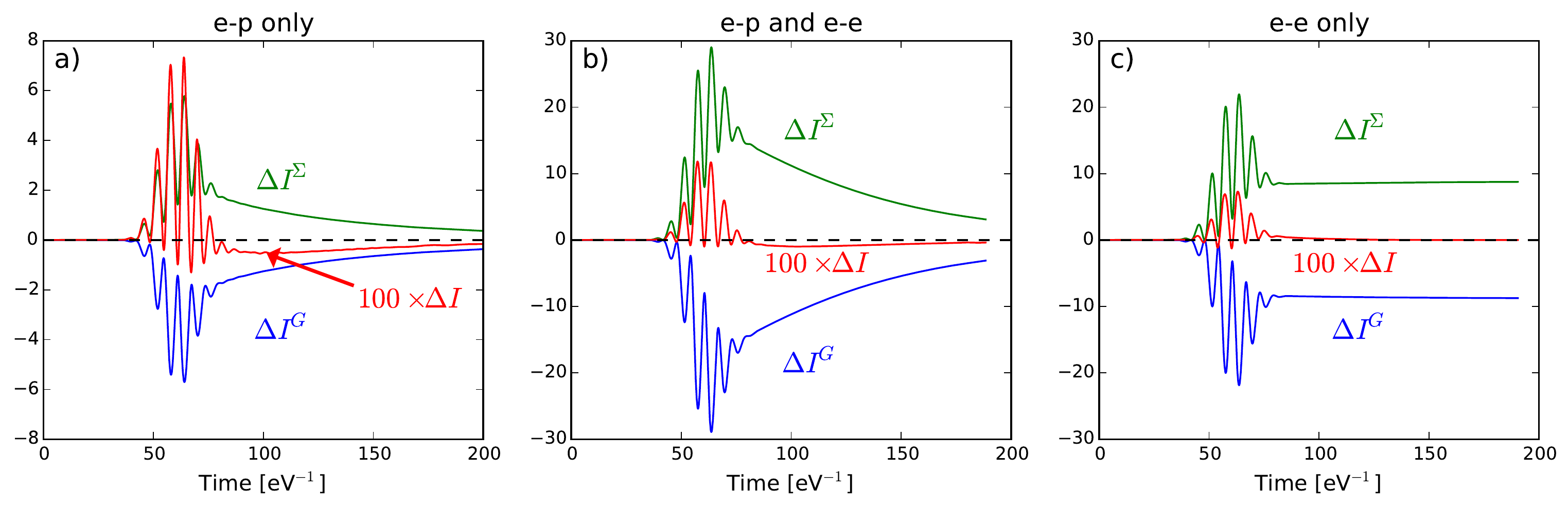}
	\caption{Time traces of the change in the momentum-summed scattering integrals $\Delta I^{\Sigma,G}(t)$ for a system with a) only e-p coupling,
	b) both e-e and e-p coupling, and c) only e-e coupling.
	Before scaling by $100$, the sum $\Delta I \equiv \Delta I^\Sigma + \Delta I^G$ (red) is nearly indistinguishable from the dashed line along the horizontal axis.
	}
	\label{fig:rhs_ksum}
\end{figure}

Fig.~\ref{fig:rhs_snapshots} shows the scattering integrals for a case with both types of scattering (e-e and e-p) at $t=108$ after the pump.
The separated terms $I^\Sigma_\kk$ and $I^G_\kk$ are both
large and appear featureless on these scales, while their sum has some definite structure.
To elucidate this further, we consider the difference from equilibrium,  $\Delta I^\Sigma_\kk$ and $\Delta I^G_\kk$,
shown in the bottom row of Fig.~\ref{fig:rhs_snapshots}.
The pump has clearly induced a difference from the equilibrium scattering integrals, giving an average time dependence to
the self-energy and Green's functions and thus resulting in a finite summed
scattering rate.
There is some momentum dependence to the terms, although the details of the momentum dependence can vary based on the
types of scattering present and their amplitudes. As time goes on, the system will reach some steady state where it has either returned to
its equilibrium state (where $\Delta I_\kk^\Sigma$ and $\Delta I_\kk^G$ also go to $0$) or where a new state is achieved, with a finite 
$\Delta I_\kk^\Sigma$ and $\Delta I_\kk^G$.

The latter can be achieved in the case of e-e scattering without any further mechanism for removing the energy from the system.
These cases can be illustrated by considering the time dependence of $\Delta I^{\Sigma,G} \equiv \sum_\kk \Delta I^{\Sigma,G}_\kk$.
Fig.~\ref{fig:rhs_ksum} shows the time traces of momentum-summed $\Delta I^\Sigma$ and $\Delta I^G$. When only e-p scattering is present,
the system behaves essentially as expected and reported previously\cite{kemper_effect_2014}
with an exponential, slow return to equilibrium together with a slow decrease of both scattering
integrals. The sum also goes to zero on a similar time scale.  It should be noted that the magnitude of $\Delta I$ is much larger
than the two contributions $\Delta I^{\Sigma,G}$, confirming again that the balance between terms controls the dynamics.
As e-e scattering is introduced, more energy is absorbed during the
pump and the scale of the changes increases, but the salient features remain the same.  Finally, when e-p scattering isn't present,
a markedly different scenario occurs. The scattering integrals remain different from their equilibrium values, but as the system relaxes,
they lose their average time dependence and balance out to give a net sum of 0 at long times.  This state reflects a ``thermalized'' electron
system, where the scattering has taken place until a balance (reflected in a Fermi distribution) is achieved.
As noted above, the balance can also be achieved with a non-thermal electron distribution, so long as the same distribution is used
in determining the self-energy.

To further illustrate the importance of cancellation between $\Delta I^\Sigma$ and $\Delta I^G$, we can estimate the decay constant
by considering the ratio 

\begin{align}
\frac{\Delta I_\kk}{\Delta n_\kk} = i \frac{\Delta I^\Sigma_\kk + \Delta I^G_\kk}{G^<(t,t)-G^<(0,0)}.
\end{align}

\noindent In the weak pumping or
non-selfconsistent limits, this is approximately (or exactly) equal to $-2\Im\Sigma(\omega=\epsilon_\kk)$.
For e-p scattering only, $\Delta I_\kk/\Delta n_\kk$ shows a reasonable agreement with the equilibrium self-energy, although some deviation
is present and in fact expected because we know that there is a contribution from the $\Delta I^G$ term in the scattering integrals. Nevertheless,
the step in the scattering rate near the phonon frequency is recognizable.  Since the scattering integral $\Delta I$ does not show any signs of a balance
before the full decay, we observe that $\Delta I_\kk/\Delta n_\kk$ remains relatively constant at early times.
When e-e scattering is included, a competition between the scattering mechanisms arises that leads to a disagreement between $1/\tau$ and
$-2\Im\Sigma_\mathrm{eq}(\omega=\epsilon_\kk))$\cite{rameau_energy_2015}. This is reflected in the scattering integrals as a deviation from simple
exponential decay, and a time dependence in $\Delta I_\kk/\Delta n_\kk$. Although the phonon window can still be observed as the energy dissipation
of the system is governed by the e-p process, the connection to the equilibrium self-energy is rapidly lost as the e-e scattering strives to achieve
a thermal balance.

\begin{figure}
\centering
	\includegraphics[width=\textwidth]{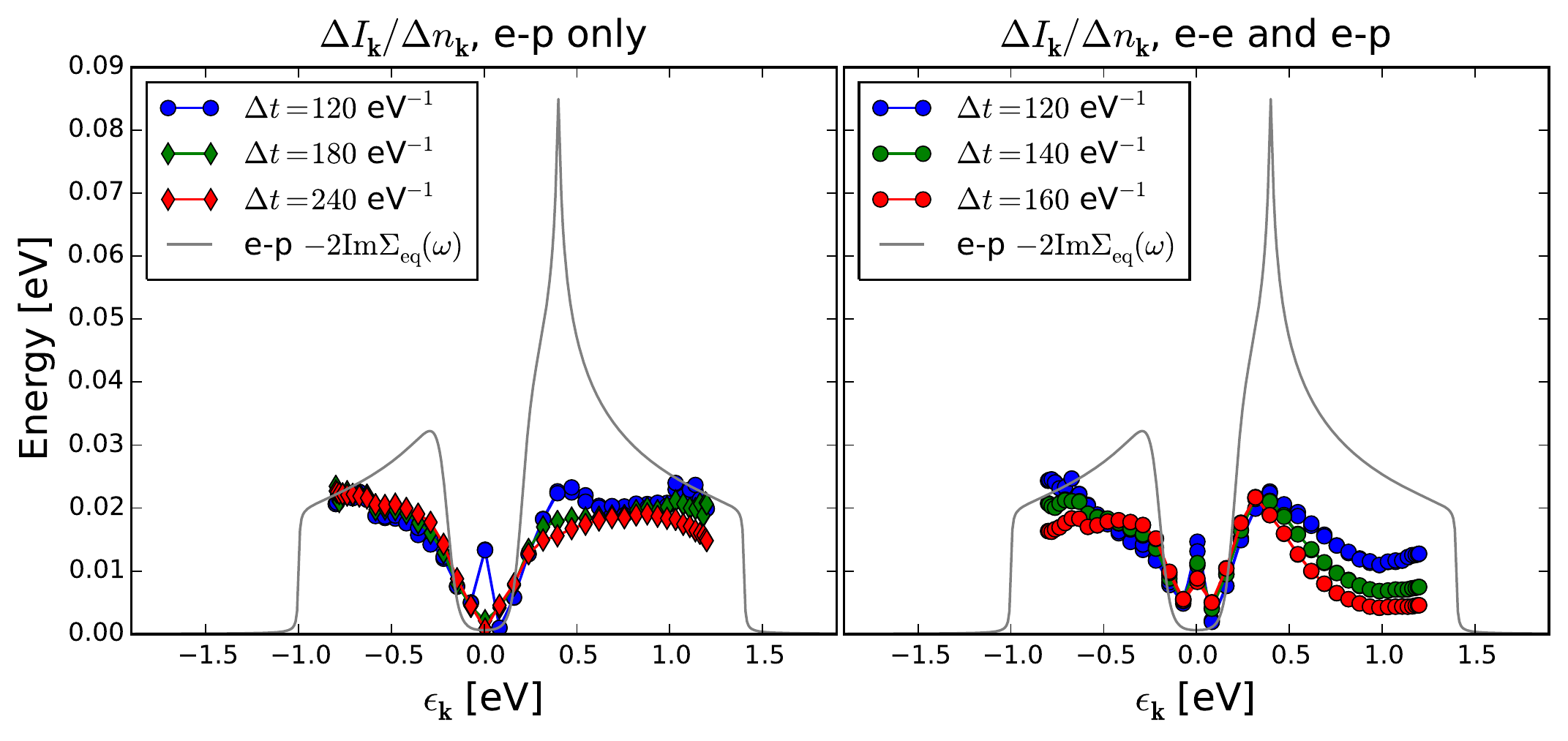}
	\caption{$\Delta I_\kk/\Delta n_\kk$ for times $\Delta t$ after $t=0$  (the pump pulse occurs at $t=60$ eV$^{-1}$). The equilibrium electron-phonon
	self-energy is shown for reference.}
	\label{fig:scattint_div_nk}
\end{figure}


\section{Discussion}
We have considered a few aspects of the equations of motion for the population of an excited system.  The major result is that a nontrivial average time
dependence of the Green's function and self-energy is needed to have any population dynamics at all, and that a steady state is reached
when there is a balance of scattering integrals after excitation rather than a full return to equilibrium of the same.  
Approximating the system with a hot electron model cannot capture the dynamics, as it was shown to lead to an exact cancellation of the scattering
integrals unless the electrons are coupled to an external bath (e.g. of phonons).
To be able to achieve this
balance, a full self-consistent treatment is necessary.  This na\"ively makes sense because the redistributed population should be accounted
for in the interactions, but stands in contrast to equilibrium physics where self-consistency often only gives a correction on top of the major feature
obtained in 1$^\mathrm{st}$ order.  For example, the kink and step in the spectral weight due to a strongly coupled boson in Migdal-Eliashberg
theory arise readily whether the bare or renormalized electron propagator is used in the calculation of the self-energy.
In the time domain, ignoring self-consistency does capture the main feature when the system is solely coupled to an external 
bath\cite{sentef_examining_2013,kemper_effect_2014}, but beyond this situation this approach is insufficient.  This leads to the conclusion that the relation between the
population scattering rate and the equilibrium self-energy  $\tau(\kk,\omega)^{-1} = -2\Im \Sigma_\mathrm{eq}(\kk,\omega)$ only holds
exactly in very limited cases for non-equilibrium experiments, although it appears to be approximately true over a much wider range of cases.

\section{Materials and Methods}
The equations of motion for the Green's function are solved using the time-stepping algorithm detailed in Ref.~\cite{stefanucci_nonequilibrium_2013}. The thermal
problem is solved on the Matsubara axis, which is followed by a forward time-stepping procedure. At each time step, the self-energy is evaluated and
the forward time step is made, followed by a correction step.  This is repeated until the Green's function at the new time is converged, and the algorithm
moves on to the next time step.

%

%
%
\vspace{6pt} 


\acknowledgments{
We acknowledge helpful conversations with H. R. Krishnamurthy. J.K.F. was supported by the U.S. Department of Energy, Office of Basic Energy Sciences, under Grant No. DE-FG02-08ER46542, and by the McDevitt bequest at Georgetown University. This research used resources of the National Energy Research Scientific Computing Center, a DOE Office of Science User Facility supported by the Office of Science of the U.S. Department of Energy under Contract No. DE-AC02-05CH11231.
}



\abbreviations{The following abbreviations are used in this manuscript:\\
\noindent 
RHS: right-hand side\\
e-p: electron-phonon\\
e-e: electron-electron
}

%

\bibliographystyle{mdpi}


\bibliography{entropy}


\end{document}